\documentclass[reprint, aps, superscriptaddress, prb, showpacs]{revtex4-1}
\usepackage{dcolumn}
\usepackage[dviout]{graphicx}
\usepackage{amsmath,amssymb}
\usepackage{bm}
\usepackage[dvipdfmx]{pict2e}
\usepackage{color}
\usepackage{url}

\newcommand{\taux}{\tau_{x}}

\newcommand{\tauz}{\tau_{z}}
\newcommand{\mux}{\mu_{x}}

\newcommand{\muz}{\mu_{z}}

\newcommand{\sigmay}{\sigma_{y}}
\newcommand{\sigmaz}{\sigma_{z}}

\newcommand{\hcosx}{\cos{\mbox{\large$\frac{k_x}{2}$}}}
\newcommand{\hcosy}{\cos{\mbox{\large$\frac{k_y}{2}$}}}
\newcommand{\hcosz}{\cos{\mbox{\large$\frac{k_z}{2}$}}}

\newcommand{\vk}{\bm{k}}
\newcommand{\vt}{\bm{t}}

\newcommand{\ket}[1]{|{#1}\rangle}

\begin{document}

\preprint{APS/123-QED}

\title {Spinless hourglass nodal-line semimetals}
\author {Ryo Takahashi}
\affiliation{
Department of Physics, Tokyo Institute of Technology, 2-12-1 Ookayama, Meguro-ku, Tokyo 152-8551, Japan\\
}
\author {Motoaki Hirayama}
\affiliation{
Department of Physics, Tokyo Institute of Technology, 2-12-1 Ookayama, Meguro-ku, Tokyo 152-8551, Japan\\
}
\affiliation{
TIES, Tokyo Institute of Technology, 2-12-1 Ookayama, Meguro-ku, Tokyo 152-8551, Japan\\
}
\affiliation{
Center For Emergent Matter Science, RIKEN, 2-1 Hirosawa, Wako, Saitama 351-0198, Japan\\
}
\author {Shuichi Murakami}
\affiliation{
Department of Physics, Tokyo Institute of Technology, 2-12-1 Ookayama, Meguro-ku, Tokyo 152-8551, Japan\\
}
\affiliation{
TIES, Tokyo Institute of Technology, 2-12-1 Ookayama, Meguro-ku, Tokyo 152-8551, Japan\\
}

\date{\today}

\begin{abstract}
Nodal-line semimetals, one of the topological semimetals, have degeneracy along nodal lines where the band 
gap is closed. In many cases, the nodal lines appear accidentally, and in such cases it is impossible to determine whether the nodal lines appear or not, only from the crystal symmetry and the electron filling. In this paper, for spinless systems, we show that in specific space groups at $4N+2$ fillings ($8N+4$ fillings including the spin degree of freedom), presence of the nodal lines is required regardless of the details of the systems. 
Here, the spinless systems refer to crystals where
the spin-orbit coupling is negligible and the spin degree of freedom can be omitted because of the SU(2) spin degeneracy. 
In this case the shape of the band structure around these nodal lines is like an hourglass, and we call this a spinless hourglass nodal-line semimetal. We construct 
a model Hamiltonian as an example and we show that it is always in the spinless hourglass nodal-line semimetal phase even when the model parameters are changed without changing the symmetries of the system. We also establish a list of all the centrosymmetric space groups, under which spinless systems always have hourglass nodal lines, and illustrate where the nodal lines are located. 
We propose that Al$_3$FeSi$_2$, whose space-group symmetry is Pbcn (No. $\bm{60}$), is one of the nodal-line semimetals arising from this mechanism.
\end{abstract}

\maketitle

\section{INTRODUCTION}
The discovery of topological insulators has triggered intensive studies on topology of the electronic band structure of crystals \cite{PhysRevLett.95.146802,PhysRevLett.95.226801,PhysRevLett.96.106802,Bernevig1757,RevModPhys.82.3045,RevModPhys.83.1057}. After the theoretical proposals, a number of materials have been experimentally shown to be topological insulators \cite{Konig766,zhang2009topological,chen2009experimental}. 
In addition, a new topological phase called a topological crystalline insulator has been proposed\cite{PhysRevLett.106.106802,hsieh2012topological,PhysRevX.6.021008,wang2016hourglass,PhysRevB.94.155148}, and  experimentally confirmed\cite{tanaka2012experimental,dziawa2012topological,xu2012observation}. In the topological crystalline insulators, topological order is realized because of crystallographic symmetries. This is one of the examples of an interplay between topology and symmetry in crystals. 

Through the researches on the topological insulators, exotic gapless phases also receive a lot of attention \cite{1367-2630-9-9-356,PhysRevB.84.235126,liu2014discovery,xu2015discovery,weng2016topological,Bradlynaaf5037}. 
Weyl semimetals and nodal-line semimetals are among such gapless phases. 
In particular, nodal-line semimetals have degeneracy along nodal lines, where the gap is closed
\cite{PhysRevB.84.235126,PhysRevB.90.115111,PhysRevB.92.081201,xie2015new,PhysRevLett.115.036806,PhysRevLett.116.127202,doi:10.7566/JPSJ.85.013708,schoop2016dirac,PhysRevB.93.205132,PhysRevB.94.195109,PhysRevB.95.041103,Bradlynaaf5037,hirayama2017topological,yu2017topological}. 
In band structures of crystals, the bands usually anticross at general $\bm{k}$ points. Therefore, 
existence of the nodal lines at general positions in $\bm{k}$ space requires special symmetries 
of the systems. 
For example, in a mirror-symmetric system, the conduction and the valence bands with opposite signs of the mirror eigenvalues do not anticross on a mirror invariant plane. If a band inversion between the conduction and the valence bands occurs at some $\bm{k}_0$ in the mirror invariant plane, 
a nodal line will appear around $\bm{k}_0$. 
We note that the nodal lines in real materials generally have dispersions
because of absence of chiral symmetry. Therefore in this paper, the word ``nodal-line semimetal'' represents a system where the nodal lines are at or near the Fermi energy.

Typically, the nodal lines can be annihilated while preserving symmetries
\cite{PhysRevB.84.235126,PhysRevB.90.115111,PhysRevB.92.081201,xie2015new,PhysRevLett.115.036806,PhysRevLett.116.127202,doi:10.7566/JPSJ.85.013708,schoop2016dirac,PhysRevB.93.205132,PhysRevB.95.041103,Bradlynaaf5037,hirayama2017topological,yu2017topological}, and therefore, symmetry alone will not tell whether the nodal lines exist or not. However,
recent investigations have demonstrated that in specific space groups at certain fillings, presence of the nodal lines is required in all realizations of crystals  \cite{PhysRevLett.115.126803,PhysRevB.94.155108,bzdusek2016nodal-nature,PhysRevB.93.155140,chen2016topological}. 
In these previous works, an hourglass-shaped band structure enforces an appearance of the nodal lines. Similar hourglass-shaped band structure 
appears as surface states in some insulating phases \cite{wang2016hourglass,PhysRevX.6.021008}.
Hence, we call this type of the nodal line an hourglass nodal line (HNL). The 
HNL corresponds to the ``orifice'' of the hourglass-shaped band structure, and at a certain filling it is enforced to be close to the 
Fermi energy. The HNLs cannot be created or annihilated while preserving all the symmetries and the electron filling.
%Apart from HNLs, another type of nodal lines may appear 
%due to high-dimensional irreducible representations, and these nodal lines are
%also enforced by symmetry, as is the case for HNLs. In the present paper we only focus 
%on HNLs. 
The HNLs have been studied mainly in spinful systems, i.e. systems with nonzero spin-orbit coupling. 
In particular, the HNLs in non-centrosymmetric spinful systems are systematically 
discussed in Ref.~\onlinecite{bzdusek2016nodal-nature}. 
On the other hand, in spinless systems, i.e. systems with negligible spin-orbit coupling, little is known about HNLs. 
Here, we note that a real crystal where
the spin-orbit coupling is negligible can be regarded as a spinless system because of the SU(2) spin degeneracy. Therefore, two bands
which are SU(2)-degenerate correspond to one band in the corresponding spinless system. Therefore, the filling factor for spinless bands is a half of that for spinful bands in real crystals.

In this paper we theoretically study the HNLs in spinless systems with inversion and time-reversal symmetries. 
%Inversion-symmetric crystals are abundant in nature and one can find various candidate materials. 
%Moreover, 
Compared to spinful systems, the nodal lines in the spinless systems have an interesting topological feature. In centrosymmetric spinless systems with time-reversal symmetry, the nodal lines 
characterized by the $\pi$ Berry phase can appear at general points in $\bm{k}$-space.
\cite{PhysRevB.93.205132}. 
%This is because in such systems, 
%the Berry phase for an arbitrary path is quantized to 0 or $\pi$. \cite{PhysRevLett.62.2747,PhysRevB.92.081201}
%In these systems, Berry phase along any loop which surrounding each 
%nodal line is also quantized
%to be $\pi$ } 
This quantized value of the $\pi$ Berry phase around the nodal lines gives robustness
to the nodal lines against perturbations which break the space-group symmetries other than the inversion symmetry. 
Furthermore, it leads to emergence of surface polarization charges, 
as discussed in Ref. \onlinecite{hirayama2017topological} on calcium. Thus the nodal lines in the spinless systems with the time-reversal and inversion symmetries have novel features among the various nodal lines which stem from space-group symmetries by various mechanisms. 
In this paper, we will show that in some centrosymmetric space groups, an appearance of the HNLs is enforced by the symmetry and the electron filling. In the spinless systems 
with such space-group symmetries, when the filling factor is $4N+2$ ($N$: integer)
excluding the spin degree of freedom (i.e. $8N+4$ including the spin degree of freedom), the gap closes along the HNLs. We find that the HNLs always run through a high-symmetry point with spinless fourfold degeneracy, which is eightfold degeneracy including the spin degree of freedom. We list all the space groups which have the HNLs enforced by the symmetries and the filling. For example, a spinless system with the $P4_2/mbc$ (No.$\bm{135}$) crystal symmetry and the time-reversal symmetry belongs to such semimetals. %\textcolor{red}{In a previous work \cite{Bradlynaaf5037}, fourfold degenerate nodal line in spinful systems with $P4_2/mbc$ symmetry are discussed. However, this nodal line comes from a four-dimensional irreducible representation, and hence it is not an HNL.} 
We construct a tight-binding model with this symmetry and show that in this system, the nodal lines enforced by the symmetries and the filling must appear and run through the $A$ point, where the states are spinless fourfold degenerate excluding the spin
degeneracy. We also show Al$_3$FeSi$_2$, which has the space group Pbcn (No. $\bm{60}$), as an example of this class of materials. 

This paper is organized as follows. In Sec.~I\hspace{-.1em}I, we show the mechanisms for emergence of the HNLs in the spinful and the spinless systems. In Sec.~I\hspace{-.1em}I\hspace{-.1em}I, we show an example of the HNLs by constructing a model Hamiltonian and study properties of the HNLs. Section I\hspace{-.1em}V is devoted to the list of all the centrosymmetric space groups with the HNLs enforced by the symmetries and the filling in the spinless systems. We also show how the nodal lines emerge in each space group. In Sec.~V, we show Al$_3$FeSi$_2$ as an example of such materials.
%\textcolor{red}{, which have space group Pbcn (No. $\bm{60}$).} 
Conclusion and discussions are given in Section V\hspace{-.1em}I. We assume presence of the time-reversal symmetry and the inversion symmetry, and absence of the spin-orbit coupling throughout the paper.

\section{Mechanism of emergence of hourglass nodal line}
In this section, we explain the mechanisms of emergence of the HNLs. 
The HNLs in the spinful systems \cite{PhysRevLett.115.126803,bzdusek2016nodal-nature,PhysRevB.93.155140,chen2016topological,PhysRevB.94.155108}, and related hourglass-shaped band structure (not limited to the nodal lines) \cite{wang2016hourglass,PhysRevX.6.021008,PhysRevB.94.155148} have been discussed previously. 
As is different from Refs.~\onlinecite{wang2016hourglass,PhysRevX.6.021008,PhysRevB.94.155148} ,
where the hourglass fermions appear as surface states, the HNLs in this paper appear in the bulk band structure. 
Here, we will show a unified description of the HNLs for the spinful and the spinless systems, and explain how 
the HNLs in the spinless systems are different from those in the spinful systems.

First, we give basic ingredients of the HNLs. Let us consider non-interacting %spinless
fermionic systems with a glide or mirror symmetry. The symmetry operation $g$ is composed of a mirror operation $\sigma$ and a translation by a vector $\vt$; $g=\{\sigma|\vt\}$. It is a glide operation when $\vt$ is a half of a translation vector, and a mirror operation when $\vt=0$. 
Hereafter we discuss the cases with the glide symmetry; those with the mirror symmetry can be treated similarly. On the glide-invariant plane in $\bm{k}$-space, the energy eigenstates are also eigenstates of the glide operator. Because of $g^2=e^{-i\vk\cdot\vt}$  in the spinless systems, the glide eigenvalues are either $e^{-i\vk\cdot\vt/2}$ or $-e^{-i\vk\cdot\vt/2}$. On
the other hand, in the spinful systems where we
include the spin degree of freedom, because of 
$g^2=-e^{-i\vk\cdot\vt}$, the glide eigenvalues take two values, $ie^{-i\vk\cdot\vt/2}$ or $-ie^{-i\vk\cdot\vt/2}$.

\begin{figure}[t]
  \centerline{\includegraphics[width=9cm,clip]{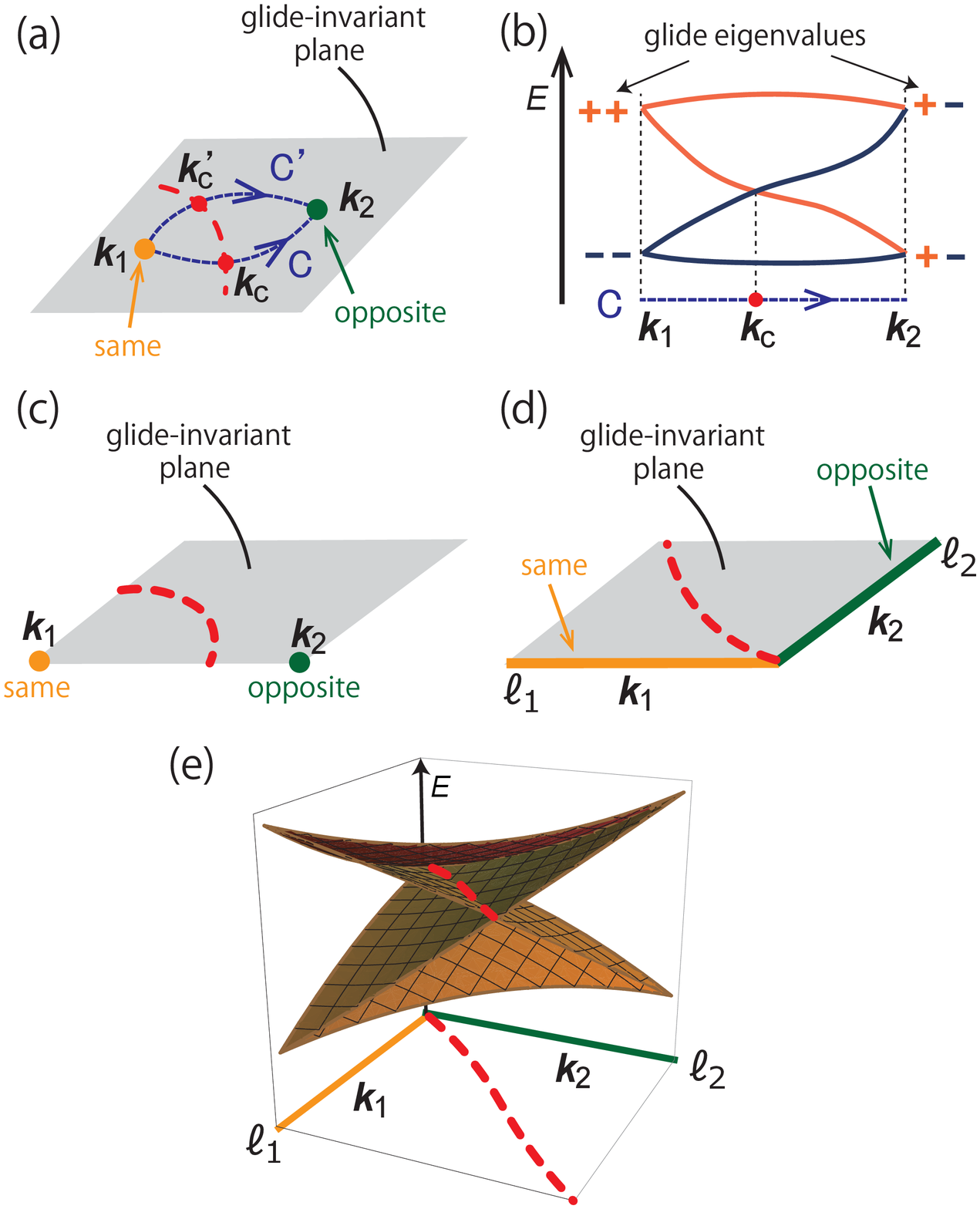}}
  \caption{(Color online) Mechanism of emergence of the HNL. 
(a) Curve $C$ which connects the two points $\bm{k}_1$ and $\bm{k}_2$ on the glide-invariant plane. 
The energy bands have twofold degeneracy at these points by symmetry. 
The glide eigenvalues of the doubly degenerate states are the same at $\bm{k}_1$ (yellow), and are of the opposite signs at $\bm{k}_2$ (green). An HNL (red dashed line) appears between $\bm{k}_1$ and $\bm{k}_2$. (b) Schematic figure of the energy bands along the curve $C$. 
Glide eigenvalues are exchanged along the curve $C$, and a degeneracy appears at $\bm{k}_{\text{C}}$. 
The signs $\pm$ denote the glide eigenvalues $\pm e^{-i\vk\cdot\vt/2}$. 
(c)(d) Schematic figures of the HNL and double degeneracy (c) in spinful systems, and (d) in spinless systems. 
(e) Schematic band structure of a spinless system with a HNL, corresponding to (d). 
}
    \label{why-HNL}
\end{figure}

We then focus on twofold degeneracies on the glide-invariant plane. The double degeneracies are classified into two cases: the glide eigenvalues ($=\pm e^{-i\vk\cdot\vt/2}$ or $=\pm ie^{-i\vk\cdot\vt/2}$) being (i) the same or (ii) of the opposite signs between the degenerate two bands. 
As we see in the following, if the two cases (i) and (ii) appear at different points on the same glide-invariant plane, the  HNL appears on the plane. 
Figure \ref{why-HNL} is the schematic figure of the HNL. In Fig.~\ref{why-HNL}(a), we assume that every energy band has twofold degeneracy at the two points $\bm{k}_1$ and $\bm{k}_2$ on the glide-invariant plane in $\bm{k}$-space. 
The reason for this twofold degeneracy depends on the system symmetries, and they
typically come from Kramers or Kramers-like degeneracy, which we explain in
detail in the respective cases. 
We also assume that 
the system symmetry enforces the emergence of 
the case (i) at $\bm{k}_1$, and that of the case (ii)  at $\bm{k}_2$. We then consider a curve $C$ which connects $\bm{k}_1$ and $\bm{k}_2$ as shown in Fig.~\ref{why-HNL}(a). Along the curve $C$, the typical band structure is shown in Fig.~\ref{why-HNL}(b). Here, because the curve $C$ is on the glide-invariant plane, the glide eigenvalue for each band remains constant along $C$. Therefore, there should be a band crossing at an intermediate point $\bm{k}_C$ on the curve $C$, at which the glide eigenvalues are exchanged. The two bands do not anticross because of the difference in the glide eigenvalues,
and an hourglass-shaped band structure appears along the curve $C$.
This discussion holds true for an arbitrary curve $C$ between $\bm{k}_1$ and $\bm{k}_2$. For example, another curve $C'$ in Fig.~\ref{why-HNL}(a) also have a band crossing point at $\bm{k'}_C$.
Therefore, a collection of $\bm{k}_C$ for the various curves $C$ forms the HNL, represented by the red dashed curve in Fig.~\ref{why-HNL}(a). Thus, the HNL corresponds to the orifice of the hourglass-shaped band structure. 
We emphasize that an appearance of the HNL is enforced by the two different types of double degeneracies at $\bm{k}_1$ and $\bm{k}_2$. When the symmetry enforces the case (i) at $\bm{k}_1$ and (ii) at $\bm{k}_2$, 
the presence of the nodal lines is required in all realizations of the energy bands at the fixed filling. If the electron filling factor is two, meaning that two bands among four in Fig.~\ref{why-HNL}(b) are filled, it 
may happen that the HNL is at or around the Fermi energy, and the system is a nodal-line semimetal, enforced both by the symmetry and the electron filling. 
In general, however, the energy is not constant along the HNL, and the system may not be 
a semimetal when the energy along the HNL disperses largely, deviating off the Fermi energy.   
We note that when applying to real electron systems with weak spin-orbit coupling, 
the actual electron filling is twice of that of spinless systems in our theory.

Types of the symmetries giving the double degeneracy required in the above scenario depends on symmetry classes of systems. We discuss here 
three typical classes of systems: (A) spinful systems without the inversion symmetry, (B) spinful systems with the inversion symmetry, and (C) spinless systems with the inversion symmetry. The first two of them are already discussed in the previous works
\cite{bzdusek2016nodal-nature,PhysRevB.93.155140,chen2016topological}. 
In these spinful cases, the double degeneracy comes from the Kramers degeneracy, 
or from the 
 ``Kramers-like'' degeneracy, which stems from a combination of some crystal symmetry and 
the time-reversal symmetry
\cite{PhysRevLett.115.126803,PhysRevB.94.155108,PhysRevX.6.021008,wang2016hourglass},
as we see below.

In (A) the spinful systems without the inversion symmetry, the time-reversal operator $\Theta$ is squared to $-1$, and this leads to the Kramers double degeneracy only at time-reversal invariant momenta (TRIM). These double degeneracies belong to the cases (i) or (ii) depending on the TRIM. That is because the glide eigenvalues $\pm ie^{-i\vk\cdot\vt/2}$ are either $\pm 1$ or $\pm i$ depending on the TRIM. If it is $\pm i$, $\Theta$ flips the sign of the glide eigenvalue, leading to the case (ii). If the glide eigenvalue is $\pm 1$, it leads to the case (i). As shown in Fig.~\ref{why-HNL}(c), the HNL appears between the two TRIM, one of which belongs to the case (i) and the other to the case (ii).

In (B) the spinful systems with the inversion symmetry, the energy bands are Kramers degenerate at all $\vk$, because $(P\Theta)^2=-1$ \cite{PhysRevLett.115.126803}, where  $P$ represents
the inversion operation. 
Therefore, the HNL with each band having no degeneracy never appears. 
%Hence, glide eigenvalues for a energy band do not have meaning generally. 
Nevertheless, if the mirror operation of the glide symmetry is off-centered, 
the Kramers degenerate states may share the same glide eigenvalue \cite{PhysRevB.92.081201,PhysRevB.94.155108,PhysRevB.95.075135}. In this case, we can discuss the 
glide eigenvalues and the HNLs for the Kramers degenerate bands. 
Though there is no systematic discussion on the HNLs in these systems, some specific examples are shown in the 
previous works \cite{PhysRevB.93.155140,chen2016topological}.

In (C) the spinless systems, $\Theta^2=+1$, and the Kramers degeneracy does not appear solely from the time-reversal symmetry. However, by combining the time-reversal symmetry $\Theta$ with non-symmorphic symmetry $G$, the new anti-unitary symmetry $\Theta G$ may cause a Kramers-like degeneracy when it squares to $-1$. Because $(\Theta G)^2=G^2=e^{-i\vk\cdot\vt}$, the Kramers-like degeneracy emerges when $\vk\cdot\vt=\pi$. This Kramers-like degeneracy appears along the 
$\Theta G$-invariant lines or on the $\Theta G$-invariant planes with $\vk\cdot\vt=\pi$. Therefore, 
double degeneracies can appear along high-symmetry 
lines in the spinless systems. In this case, the HNL can appear between the two high-symmetry lines as shown in Fig.~\ref{why-HNL}(d)(e).

\begin{figure}[t]
  \centerline{\includegraphics[width=8cm,clip]{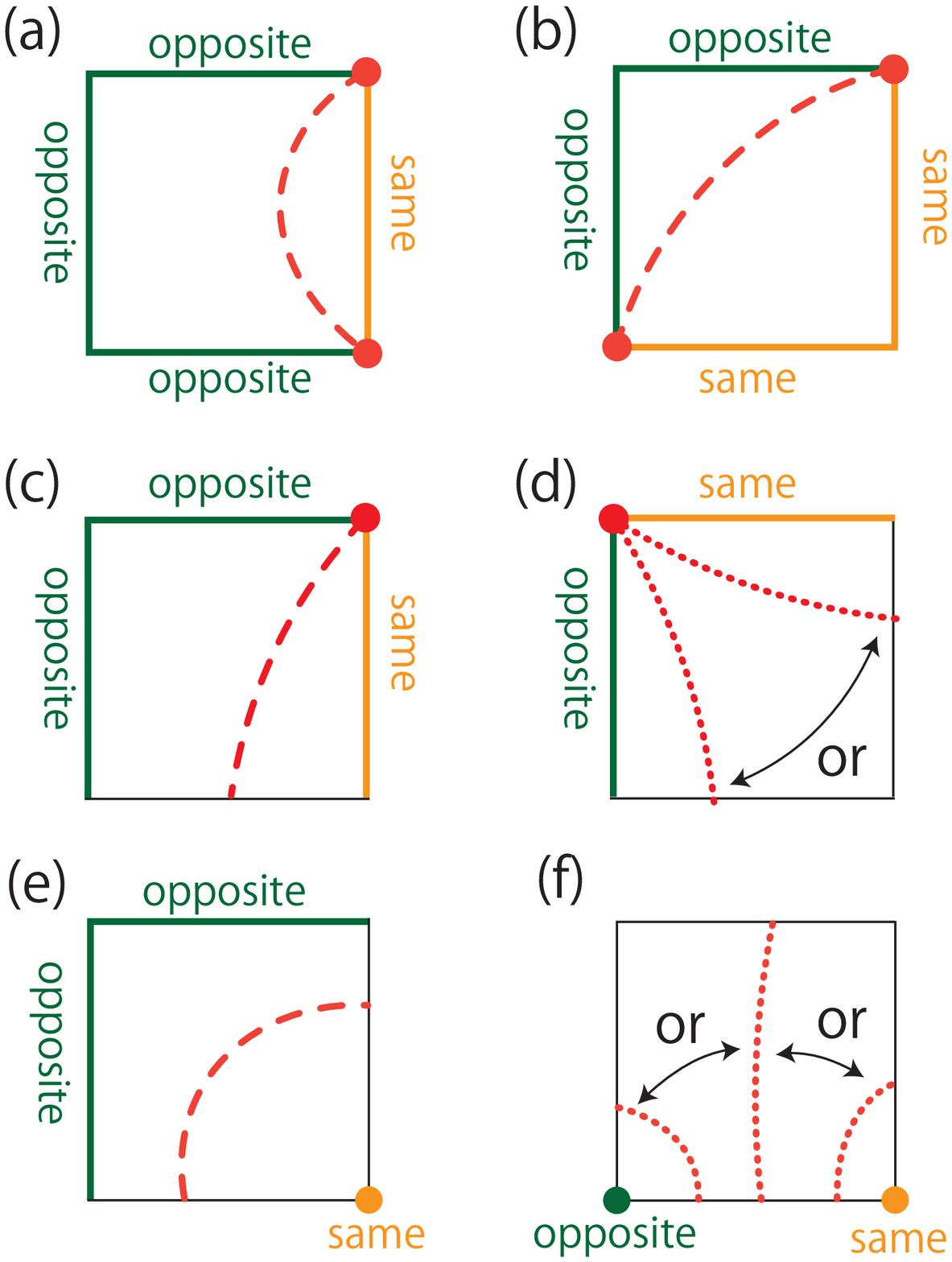}}
  \caption{(Color online) Six representative examples of the HNLs on a glide-invariant plane. Only a quarter of a two-dimensional slice of the three-dimensional Brillouin zone is shown, and the four corners represent different time-reversal invariant momenta (TRIM). The green and the yellow heavy lines and points mean twofold degeneracy with the opposite or the same glide eigenvalues, respectively. In (a)-(d), fourfold-degenerate points (red points) appear at the intersections of the green and the yellow lines. On the black thin lines, there is no band degeneracy coming from symmetries. The dashed or the dotted red lines are the HNLs, and they should appear between the green and the yellow lines. In (a)-(c), connectivity of the HNLs is determined as the dashed lines, whereas in (d) there are two possibilities for the connectivity of the HNL, shown as the dotted lines. (e)-(f) the HNLs appear between a point and lines, or between two points. In (f), there are three possibilities for the connectivity of the HNL, shown as dotted lines.
}
    \label{why-nodal-4}
\end{figure}

Thus, the appearance of the HNL in the spinless systems is guaranteed by the two doubly-degenerate high-symmetry lines, one with the same glide eigenvalues, and the other with the opposite glide eigenvalues. By considering various high-symmetry lines on each glide-invariant plane, positions of the nodal lines can be determined. Four representative examples are shown in Figs.~\ref{why-nodal-4}(a)-(d). In Figs.~\ref{why-nodal-4}(a)-(d), the high-symmetry lines with the same glide eigenvalues are shown in yellow, and those with the glide eigenvalues of the opposite signs are shown in green. The four corners in each panel represent different TRIM. In (a)-(d), the fourfold-degenerate points (red points) appear at the intersections between the green and the yellow lines. The HNLs appear between the green and the yellow lines, and start from the fourfold-degenerate points. In (a)-(c), connectivity of the nodal lines is determined as dashed lines, whereas in (d) there are two possibilities for the nodal lines, shown as the dotted lines. 
So far, we have discussed the HNLs appearing between the two high-symmetry lines. Other types of the HNLs appear between two high-symmetry points, or between a high-symmetry point and a high-symmetry line, as shown in Figs.~\ref{why-nodal-4}(e) and (f). In spinful systems, many cases of HNLs \cite{bzdusek2016nodal-nature}
belong to these types, with the high-symmetry points being the TRIM.

In many cases, space-group symmetry alone does not guarantee an appearance of the HNLs. 
That is because there are usually several two-dimensional irreducible representations at the 
same $\bm{k}$ point, and some of them belong to (i) and the rest belong to (ii).  For example, as we will see later in Sec.~I\hspace{-.1em}I\hspace{-.1em}I, there are four possibilities of double degeneracy (i.e. two-dimensional irreducible representations) at the Z point in the space group No.$\bm{135}$. Two of them correspond to the case (i), i.e. doublets with the combinations $(++)$ or $(--)$ for $G_x$ and $G_y$ eigenvalues, and the other two correspond to the case (ii), i.e. a doublet with $(+-)$. In this case, presence or absence of the HNLs depends on details of the systems. 
%Which case realized depends on eigenvalues of $\pi$-rotation around $z$-axis.
Nevertheless, we show in the following that in some space groups, existence of the HNL is guaranteed solely by the space-group symmetry. These space groups have limited combinations of the glide eigenvalues in the irreducible representations. In these space groups, on two lines $\ell_1$ and $\ell_2$ in $\bm{k}$-space, the glide eigenvalues are always as shown in Fig.~\ref{why-HNL}(d). That makes the HNL to be enforced solely by symmetry. From the list of irreducible representations of the space groups\cite{Aroyo:xo5013}, we can determine what kind of band degeneracy appears on high-symmetry lines, and whether the space groups host the HNLs or not.

\begin{figure}[t]
  \centerline{\includegraphics[width=6cm,clip]{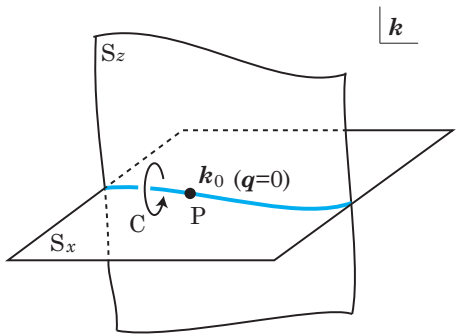}}
  \caption{(Color online) Hourglass nodal line and its Berry phase from the 
two-band effective model. The blue curve is the nodal line on the mirror (glide)
plane S$_x$, and the Berry phase along the path C around the nodal line is equal to $\pi$ (mod $2\pi$).
}
    \label{Berry-phase}
\end{figure}

The Berry phase around each of the HNLs demonstrated in the present section is always equal to $\pi$. In this sense they are topological and are robust against perturbations which preserve the time-reversal and the inversion symmetries. This is shown in the following way, by constructing an effective model for the 
two bands which cross at the nodal line. 
On the nodal line, 
we pick up an arbitrary point P with $\bm{k}=\bm{k}_0$ which is not a high-symmetry point, 
and construct a two-band effective model $H(\bm{k})$ which is valid near $\bm{k}_0$. 
For notational simplicity we take the $z$ axis to be perpendicular to the mirror (glide) plane. 
Then $\bm{k}_0$ is on the mirror (glide) invariant plane, meaning that $(0,0,2k_{0,z})$ is 
a reciprocal lattice vector. 
With this setup, we follow the method similar to Ref.~\onlinecite{okugawa17}. 
Here we take the basis set to be the two bands, which have the opposite signs of the 
mirror or glide eigenvalues  on the mirror (glide) plane. 
We expand $H(\bm{k})$ in terms of the Pauli matrices, and discard the part proportional to 
the unit matrix because it does not affect the Berry phase here. 
Then, from the time-reversal and the inversion symmetries, the effective model 
has only two Pauli matrices $\{\sigma_x,\sigma_z\}$ or $\{\sigma_y,\sigma_z\}$, among the three matrices $\sigma_x$, $\sigma_y$ and $\sigma_z$, as shown in Ref.~\onlinecite{okugawa17}. 
By some gauge transformation, one can always write the Hamiltonian as 
\begin{equation}
H(\bm{k})=a_x(\bm{k})\sigma_x+a_z(\bm{k})\sigma_z,
\end{equation}
where $a_x$ and $a_z$ are real. 
Because the mirror (or glide) operator is proportional to $\sigma_z$ due to the 
different signs of the mirror (or glide) eigenvalues, the above Hamiltonian satisfies
$\sigma_zH(k_x,k_y,k_z)\sigma_z=H(k_x,k_y,-k_z)$. 
Therefore if we take $\bm{q}\equiv \bm{k}-\bm{k}_0$, we can write this symmetry constraint
as 
$\sigma_zH(q_x,q_y,q_z)\sigma_z=H(q_x,q_y,-q_z)$ when the Hamiltonian is expressed as a function of $\bm{q}$, because $\bm{k}_0$ is on the mirror (glide) invariant plane. Then by expanding the Hamiltonian to the lowest order in $\bm{q}$ we
conclude $a_x=q_zf_x(q_x,q_y,q_z^2)$ and  $a_z=f_z(q_x,q_y,q_z^2)$, where $f_x$ and $f_z$ are
analytic functions. By definition, $f_z$ vanishes at the point P ($\bm{q}=0$), and therefore an equation
$f_z=0$ determines a curved surface S$_z$ which passes through the point P. On the other hand,
an equation $a_x=0$ is satisfied on the mirror (glide) plane S$_x$, which also passes through the point P.
The intersection between S$_x$ and S$_z$ is the nodal line, where the two eigenenergies are
degenerate. We show it schematically in Fig.~\ref{Berry-phase}. Hence, the Berry phase around this nodal line is 
calculated easily, and it turns out to be a half of the phase change of $a_x +ia_z$ around the nodal line. It is equal to $\pi$ (modulo 2$\pi$). 
This $\pi$ Berry phase of each nodal line directly affects the $\mathbb{Z}_2$ topological number 
characterizing the nodal-line semimetals \cite{PhysRevLett.115.036806,okugawa17}. Namely, if the HNLs in the present paper penetrate the half of  the Brillouin zone (in the definition of the $\mathbb{Z}_2$ topological number)  odd times, the corresponding $\mathbb{Z}_2$ topological number is nontrivial.
This applies to all the HNLs in the present paper.

The $\mathbb{Z}_2$ topological number for topological nodal lines and 
the $\pi$ Berry phase is important with regards to 
topological insulators as well \cite{PhysRevLett.115.036806,PhysRevB.85.115105}. 
It implies that for a real 
crystal having a nodal line with vanishing spin-orbit coupling, by adding
a spin-orbit term to open a gap along the nodal lines, the system becomes  
a topological insulator if the time-reversal symmetry is preserved. 
It is because the $\mathbb{Z}_2$ topological number  determining existence of  
an odd number of line nodes piercing a plane in spinless systems is also the topological number
for weak topological insulators for that plane after introducing spin-orbit-coupling terms.

\section{Model calculation}
In this section, we introduce a spinless tight-binding model which has two types of the HNLs, one 
of which is enforced by the space-group symmetry and the electron filling. 
We show an appearance of the HNLs in this model, and see parameter dependence to show
that some of the HNLs are enforced by the symmetry.

\subsection{Tight-binding model with the $P4_2/mbc$ symmetry and its band structure}
\begin{figure}[t]
  \centerline{\includegraphics[width=8.5cm,clip]{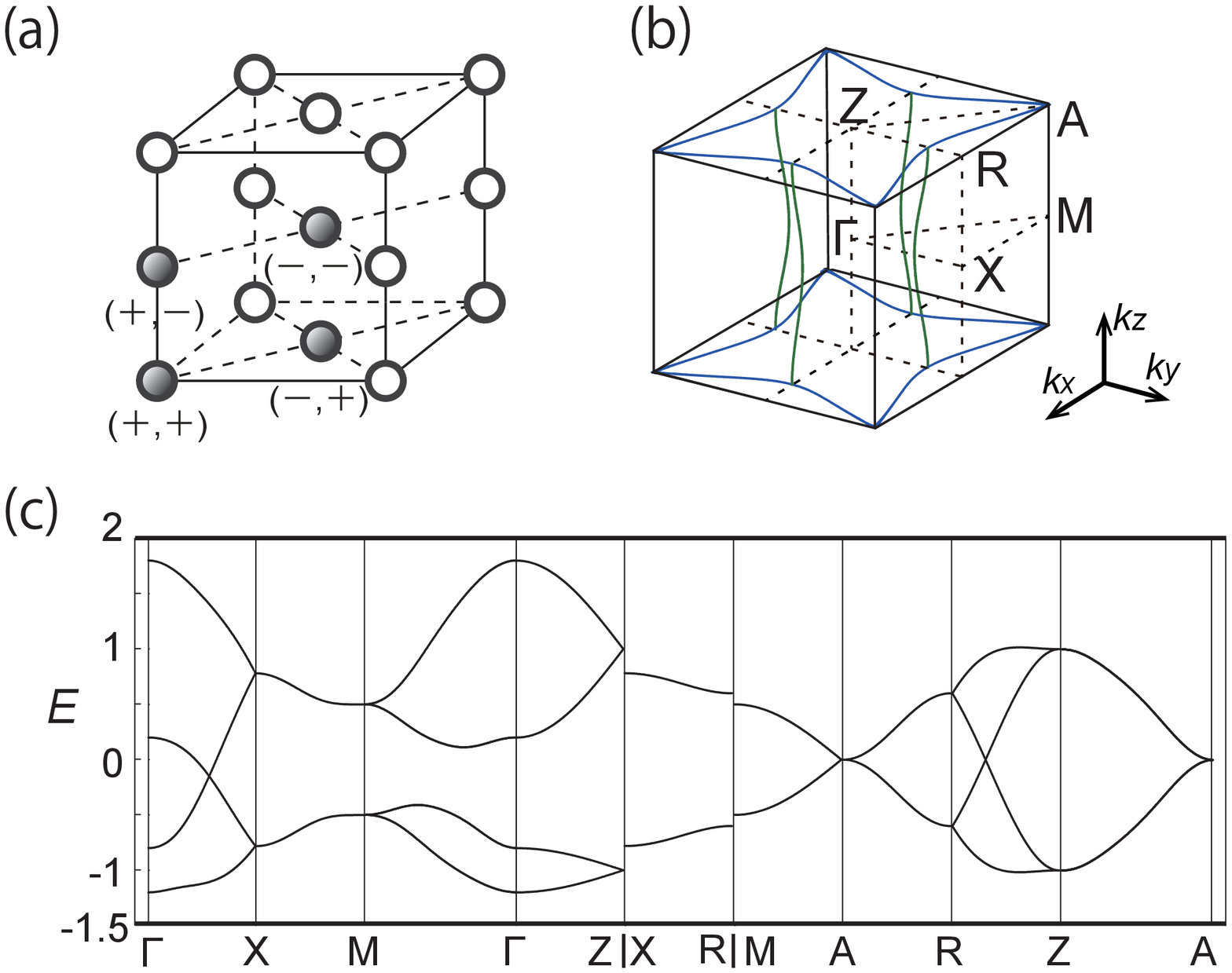}}
  \caption{(Color online) (a) Tetragonal crystal structure of our model. The four gray spheres denote the unit cell. (b) HNLs of the model (1), shown in the Brillouin zone enclosed by the planes $k_i=\pm\pi$ ($i=x,y,z$). There are two types of the HNLs, one in blue and the other in green. The HNLs on the $k_z=\pi$ plane, drawn in blue, are enforced solely by the symmetry, while the HNLs on the planes $k_x=0$ and $k_y=0$ (green curves) are not. (c) Band structure of the model (1). The parameters are $t_{xy}=1.0,\ t_z=0.5$, and $t_a=t_b=t_c=0.3$.
}
    \label{kessyou1}
\end{figure}
\renewcommand{\arraystretch}{1.3}
\begin{table}
\caption{Some symmetry operations in the $P4_2/mbc$ symmetry. The glide/screw operations
are written as combinations of a mirror reflection/rotation and a translation, following the notations in Refs. \onlinecite{ITA,burns}. In Ref. \onlinecite{ITA}, $G_x$ is a $b$-glide, $G_y$ is a $a$-glide, $G_{x+y}$ is a $c$-glide and $G_{x-y}$ is an $n$-glide. 
The translations contained in the glide operations are along the mirror plane, except for 
the $n$-glide, $G_{x-y}$, as noted in Sec.~5.2.2. in Ref.~\onlinecite{burns}. }
\label{symmetry-tab}
\tabcolsep8pt\begin{tabular}{cl}
\hline
\hline
$M_z$ & $(x,y,z) \to (x,y,-z)$ \\
$G_x$ & $(x,y,z) \to (\frac{1}{2}-x,y,z)+(0,\frac{1}{2},0)$ \\
$G_y$ & $(x,y,z) \to (x,\frac{1}{2}-y,z)+(\frac{1}{2},0,0)$ \\
$G_{x-y}$ & $(x,y,z) \to (y,x,z)+(\frac{1}{2},\frac{1}{2},\frac{1}{2})$ \\
$G_{x+y}$ & $(x,y,z) \to (\frac{1}{2}-y,\frac{1}{2}-x,z)+(0,0,\frac{1}{2})$ \\
$S_{2x}$ & $(x,y,z) \to (x,\frac{1}{2}-y,-z)+(\frac{1}{2},0,0)$ \\
$S_{2y}$ & $(x,y,z) \to (\frac{1}{2}-x,y,-z)+(0,\frac{1}{2},0)$ \\
$S_{4z}$ & $(x,y,z) \to (-y,x,z)+(0,0,\frac{1}{2})$ \\
$P$ & $(x,y,z) \to (-x,-y,-z)$ \\
\hline
\hline
\end{tabular}
\end{table}
\renewcommand{\arraystretch}{1}

We construct a tight-binding model whose space group is $P4_2/mbc$ (No.$\bm{135}$), having the mirror symmetry $M_z$, the four glide symmetries $G_x,G_y,G_{x\pm y}$ and the inversion symmetry $P$, as listed in Table I. 
This model is a spinless counterpart to the
model proposed in Ref.~\onlinecite{PhysRevLett.116.186402} of a double Dirac semimetal in the space group No.$\bm{135}$.
The difference between the two models is whether the spin-orbit coupling is zero or nonzero, 
and it makes a large difference in the band structure; the spinful model has a double Dirac node, and the spinless one in our paper has the HNLs. 

%%aaaaaaaaaaaaaaa

The model is constructed on a tetragonal lattice with four sub-lattices in the unit cell labeled by $(\tauz,\muz)=(\pm 1,\pm 1)$ (Fig.~\ref{kessyou1}(a)), and each sub-lattice has one state. Our tight-binding Hamiltonian is defined as follows: 
\begin{align}
\mathcal{H}(\vk)
&=t_{xy}\taux\hcosx\hcosy+t_z\mux\hcosz \notag \\
&\ +t_a\taux\mux\hcosx\hcosy\hcosz\notag \\
&\ +t_b\muz\left(\cos k_x- \cos k_y\right) + t_c\tauz\muz\sin k_x\sin k_y,
\end{align}
where $t_{xy},t_{z},t_a,t_b,$ and $t_c$ are real hopping amplitudes, and $\tau_i,\mu_i$ are the Pauli matrices acting on the sublattice degree of freedom. Here, the lattice constants along the three axes are set as unity for simplicity. Its band structure is shown in Fig.~\ref{kessyou1}(c), where we set the parameters as $t_{xy}=1.0,\ t_z=0.5$ and $t_a=t_b=t_c=0.3$. We can see some degeneracy along the $\Gamma$-X and R-Z lines between the second and the third bands in the band structure. These degeneracies are not limited to isolated points in $\bm{k}$-space but are a part of HNLs. The HNLs of this model are shown in Fig.~\ref{kessyou1}(b). This model has two types of HNLs; one is on the $k_z=\pi$ plane and the other is on $k_x=0$ and on $k_y=0$. The HNLs on the mirror-invariant plane $k_z=\pi$ (blue curves)
are enforced by the system symmetry, whereas 
those on the $k_x=0$ and $k_y=0$ planes (green curves) are not, as we show in the following.

Before discussing the origin of the HNLs, we explain double degeneracies in the band structure of the model. Because of the symmetry, the energy bands have twofold degeneracy along several high-symmetry lines such as X-M, X-R, M-A, A-R, and Z-A lines, as determined from the irreducible representations of the space group No.$\bm{135}$ \cite{Aroyo:xo5013}. For example, there is only one two-dimensional irreducible representation along the Z-A line. It gives degeneracy between two states with the mirror eigenvalue of $M_z$ being of the opposite signs. 
This double degeneracy can also be understood as the Kramers-like degeneracy due to the 
$\Theta S_{2y}$ symmetry. Because $(\Theta S_{2y})^2=e^{-ik_y}$, the energy bands are Kramers-like-degenerate at all $\bm{k}$ on the $k_y=\pi$ plane. This explains the double degeneracy along the X-M, X-R, M-A and A-R lines. 
The double degeneracy along the Z-A line can also be understood as the Kramers-like degeneracy due to the $\Theta G_{x-y}$ symmetry. Because $(\Theta G_{x-y})^2=e^{-i(k_x+k_y+k_z)}$, the energy bands are Kramers-like degenerate at all $\bm{k}$ along the Z-A line.

We now explain the origin of the HNLs of the model. First, let us focus on the $k_z=\pi$ plane, which is invariant under the mirror operation $M_z$. All the energy bands have twofold degeneracy along the A-R line and the Z-A line on the $k_z=\pi$ plane. Along the Z-A line, two Bloch states with opposite $M_z$ eigenvalues are degenerate, corresponding to the case (ii). 
That is because the $\Theta G_{x-y}$ symmetry, which causes the Kramers-like double degeneracy along the Z-A line, anti-commutes with $M_z$ on $k_z=\pi$: 
\begin{align}
(\Theta G_{x-y}) M_z = e^{-ik_z} M_z (\Theta G_{x-y}).
\end{align}On the other hand, along the A-R line, two Bloch states with the same mirror $M_z$ eigenvalue are degenerate, i.e. the case (i) emerges. 
That is because the $\Theta S_{2y}$ symmetry, which causes the Kramers-like double degeneracy along the A-R line, commutes with the $M_z$ symmetry: 
\begin{align}
(\Theta S_{2y}) M_z = M_z (\Theta S_{2y}).
\end{align}Between the Z-A line and the A-R line, the mirror $M_z$ eigenvalues are exchanged, and an HNL appears between them as shown in Fig.~\ref{why-HNL}(d). We emphasize that an appearance of this HNL is enforced solely by the symmetry and the filling. 
Even when we include all the irreducible representations allowed by symmetry, the $M_z$ eigenvalues of the doubly degenerate states are always of the opposite signs 
(case (ii)) along the Z-A line and are the same (case (i)) along the A-R line;
no other possibility of band degeneracy is allowed along the Z-A line and along the A-R line,
because of the symmetry.

%The origin of the double degeneracy is the non-commutative nature of the non-symmorphic glide symmetry. For example, along the Z-A line, the glide operation $G_{x+y}$ anti-commutes with the mirror operation $M_z$, giving double degeneracy between states with opposite $M_z$ eigenvalues. On the other hand, along the A-R line, the glide operation $G_y$ anti-commutes with the product of time-reversal operation $\Theta$ and inversion operation $P$, giving double degeneracy between states with same $M_z$ eigenvalues. 

Next, we focus on the $k_x=0$ plane, which is invariant under the glide operation $G_x$. The energy bands have twofold degeneracy along the X-R line and at the Z point. Along the X-R line, two Bloch states with the opposite $G_x$ eigenvalues are degenerate. 
Because the $\Theta S_{2y}$ symmetry commutes with the $G_x$ symmetry, and the  eigenvalues of $G_x$ are $\pm e^{-ik_y}$, which is $\pm i$ along $k_y=\pi$, we  get
\begin{align}
G_x(\Theta S_{2y})\ket{G_x=+i}
&=(\Theta S_{2y})(G_x\ket{G_x=+i}) \notag \\
&=(\Theta S_{2y})(+i\ket{G_x=+i}) \notag \\
&=-i(\Theta S_{2y})\ket{G_x=+i}.
\end{align}
On the other hand, at the Z point, two Bloch states with the same $G_x$ eigenvalue are degenerated. 
At the Z point, eigenvalues of the $G_x$ are $\pm 1$, and the $\Theta G_{x-y}$ symmetry commutes with the $G_x$ symmetry because $C_{2z}=1$ in this model. Therefore, the two states in the Kramers-like degeneracy by the $\Theta G_{x-y}$ symmetry share the same eigenvalues of the $G_x$. 
\begin{align}
G_x (\Theta G_{x-y})\ket{G_x=+1}
&=e^{ik_x}(\Theta G_{x-y})G_x C_{2z}\ket{G_x=+1} \notag \\
&=(\Theta G_{x-y})(G_x\ket{G_x=+1}) \notag \\
&=(\Theta G_{x-y})\ket{G_x=+1}.
\end{align}
Therefore, between the X-R line and the Z point, the glide eigenvalues are exchanged, and an HNL must appear between them, as is similar to Fig.~\ref{why-nodal-4}(e). Although the appearance of this HNL is guaranteed by the double degeneracy along the X-R line and at the Z point, this HNL is not enforced by the symmetry in general. That is because there are several irreducible representations at the Z point. %\textcolor{red}{depending on the eigenvalue of $C_{2z}$} 
In some of them, the doubly degenerate states have the same $G_x$ eigenvalues, 
while in others, they have the opposite signs of  the $G_x$ eigenvalues. In the latter case
HNLs do not exist between the X-R line and the Z point. 
Thus one cannot conclude existence of HNLs from the symmetry alone, and presence or
absence of HNLs depends on systems.
Thus this HNL between the X-R line and the Z point is not enforced solely by the symmetry.

%\subsection{\textcolor{red}{Parameter independence} of hourglass nodal lines}
\subsection{Parameter dependence of the hourglass nodal lines}
\begin{figure}[t]
  \centerline{\includegraphics[width=8cm,clip]{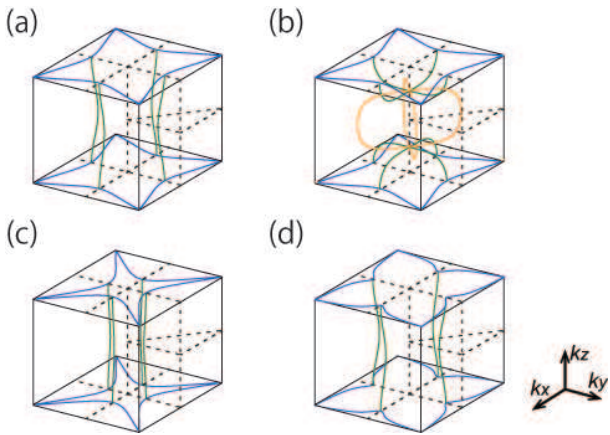}}
  \caption{(Color online)  Positions of the nodal lines of the model (1) for various values of the parameters. 
In (a), the values are $t_{xy}=0,\ t_z=0.5,\ t_a=t_b=t_c=0.3$.
In (b), only $t_a$ is changed to $t_a=1.5$. 
In (c), only $t_b$ is changed to $t_b=1.5$. 
In (d), only $t_c$ is changed to $t_c=1.5$. 
The colored lines are the nodal lines. Among the nodal lines, only the blue lines are the HNLs enforced by the symmetry.
In (b), additional nodal lines appear on the $k_x=\pm k_y$ planes, shown in yellow. 
They are not enforced by symmetry. 
}\label{robust}
\end{figure}

Thus, we have shown that emergence of the HNL on the $k_z=\pi$ plane is enforced 
by the symmetry. That means that this HNL should appear independently of the values of the parameters of the Hamiltonian. In our model, we checked this by changing the three parameters $t_a,t_b,t_c$ of the model Hamiltonian. In Fig.~\ref{robust}, we show the positions of the nodal lines for various values of the parameters $t_a,t_b$ and $t_c$. We can see that the nodal lines on the $k_z=\pi$ move, but do not disappear when the parameters change. 

Similarly, the HNLs on the $k_x=0$ plane and on the $k_y=0$ plane do not disappear against a change of the parameters. It is because in this particular model, the glide eigenvalues are always exchanged between the X-R line and the Z point.
%, \textcolor{red}{because of $\pi$-rotation around $z$-axis ($C_{2z}$) is represented as $C_{2z}=1$. 
%The HNLs are not enforced by symmetry, but representation of symmetry.} 
Therefore, the HNLs may move, but never disappear.

However, there is a difference between the HNLs on the $k_z=\pi$ plane and the HNLs on the $k_x=0$ and $k_y=0$ planes. The former is enforced by the symmetry, but the latter is not, as
shown in the previous subsection.
Therefore, in general systems with the same symmetry, the former always exists, but the latter may not exist.

\section{Generalization to other space groups}
In this section, we generalize our results to other space groups that have inversion symmetry and glide (or mirror) symmetries. Time-reversal and inversion symmetries are assumed throughout our theory. Nodal lines in the spinless systems
under these symmetries have a special feature, unlike the spinful systems. 
In the centrosymmetric spinless systems with the time-reversal symmetry, nodal lines  at general positions in $\bm{k}$-space 
are always
characterized by $\pi$ Berry phase.
In such systems, 
the Berry phase for an arbitrary path is quantized to 0 or $\pi$. \cite{PhysRevLett.62.2747,PhysRevB.92.081201}
In these systems, the Berry phase along any loop  surrounding a
nodal line is quantized
to be $\pi$. 
That makes the HNLs to be robust against perturbation which preserves the inversion symmetry but breaks other crystallographic symmetries. 

In the following, we list all the space groups having the HNLs enforced solely by the symmetry. It is a cumbersome task to find out the positions of the HNLs enforced by the symmetry for all the space groups, because there are varieties of $\bm{k}$-points and irreducible representations. This task can be facilitated in the following way. 
The hourglass-shaped band structure with the HNLs (Fig.~\ref{why-HNL}(b)) consists of four spinless energy bands, excluding the spin degree of freedom. If we include the spin degree of freedom, it consists of eight energy bands. 
This means that the minimal filling to realize a band insulator should be equal or larger than eight in candidate space-group symmetries. Here, we can use the theory by Watanabe, Po, Zaletel and Vishwanath, which recently showed a lower bound of the minimal insulating filling for 230 space groups \cite{PhysRevLett.117.096404}. From their work \cite{PhysRevLett.117.096404}, the following 17 space groups, having the inversion symmetry, are candidate space groups with  HNLs enforced by symmetry and filling: $\bm{52}$, $\mathbf{54}$, $\mathbf{56}$, $\mathbf{57}$, $\mathbf{60}$, $\mathbf{61}$, $\mathbf{62}$, $\mathbf{73}$, $\mathbf{130}$, $\mathbf{133}$, $\mathbf{135}$, $\mathbf{138}$, $\mathbf{142}$, $\mathbf{205}$, $\mathbf{206}$, $\mathbf{228}$ and $\mathbf{230}$. By examining all these space groups as explained in the following Secs.~\ref{sec:ortho}-\ref{sec:nonpri}, we find that the following 11 space groups have the HNLs enforced by the symmetry and the filling: $\bm{52}$, $\mathbf{54}$, $\mathbf{56}$, $\mathbf{57}$, $\mathbf{60}$, $\mathbf{61}$, $\mathbf{62}$, $\mathbf{130}$, $\mathbf{135}$, $\mathbf{138}$, $\mathbf{205}$. Here, as we stated in the introduction, the time-reversal symmetry is assumed. 
%General systems with other space group symmetries have 8-fold degenerated points, but may not have HNLs. 
In the following, we show the positions of the symmetry-enforced HNLs for these space groups.

\subsection{Orthorhombic primitive space groups}
\label{sec:ortho}
\begin{figure*}[t]
\centerline{\includegraphics[width=16cm,clip]{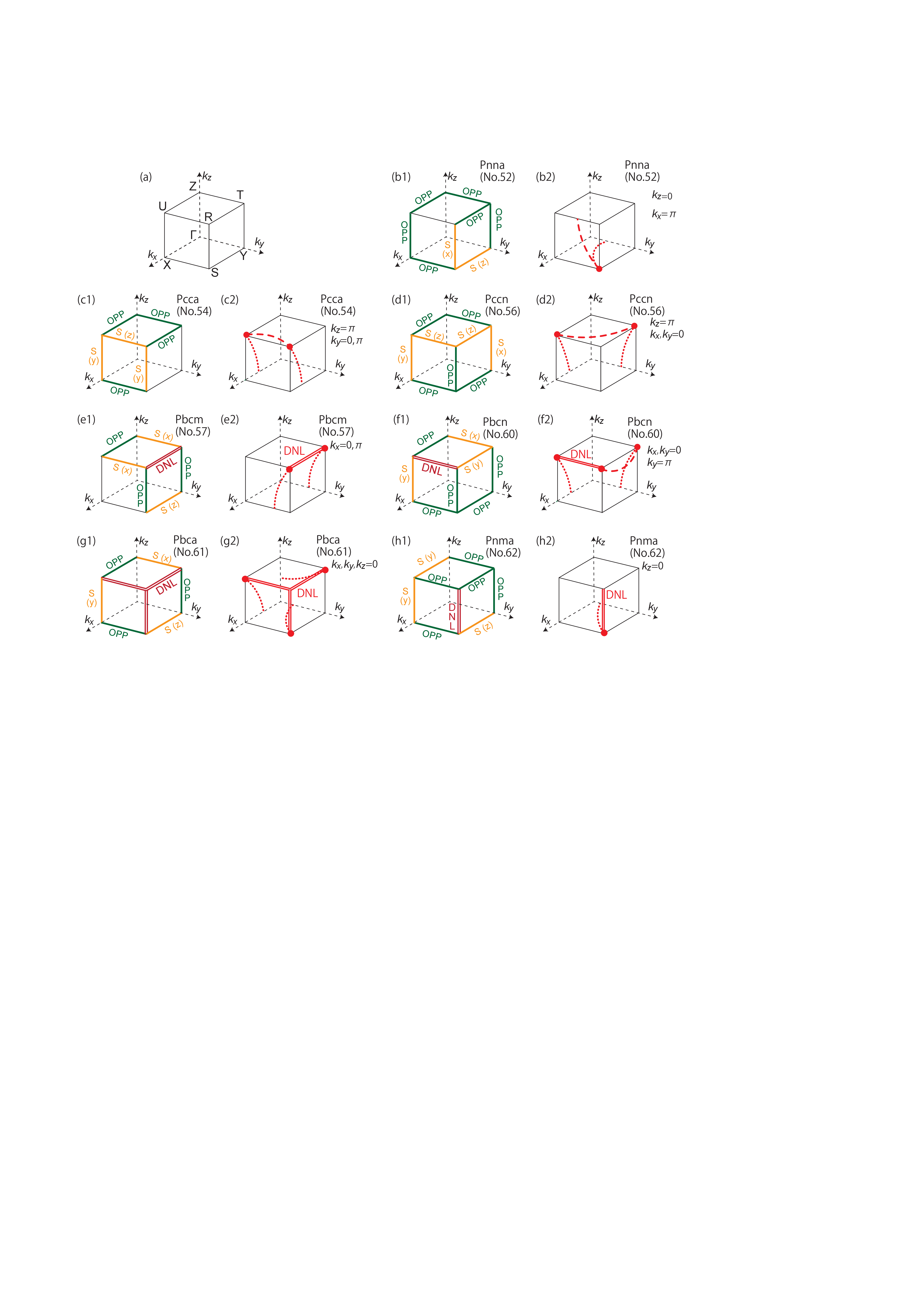}}
  \caption{HNLs enforced by the symmetry in the orthorhombic space groups No.~$\bm{52}$, $\bm{54}$, $\bm{56}$, $\bm{57}$, $\bm{60}$, $\bm{61}$ and $\bm{62}$. (a) Brillouin zone and the symmetry labels for the orthorhombic primitive lattice. (b1)-(h1) Degeneracies along the high-symmetry lines and their glide eigenvalues, which correspond to high-dimensional irreducible representations. The yellow and the green heavy lines mean twofold degeneracy with the same and the opposite glide eigenvalues, labeled with S $(x,y,z)$ and OPP, respectively. S $(x)$ means that the glide eigenvalues with the glide plane perpendicular to the $x$ direction are the same between the degenerated two bands, and S$(y)$ and S$(z)$ are defined similarly. The red double lines mean a double nodal line (DNL) having fourfold degeneracy. (b2)-(h2) HNLs in these space groups. The dashed lines (red) and the dotted lines (red) are the HNLs. Connectivity of the dotted nodal lines has more than one possibilities from the symmetry alone (see Fig.~\ref{why-nodal-4}(d)), and in the figures we only show one possibility for the positions of the HNLs, as the dotted lines.}
\label{ortho}
\end{figure*}
In this section, we discuss systems with the following 7 orthorhombic space groups: No.$\bm{52}$, $\bm{54}$, $\bm{56}$, $\bm{57}$, $\bm{60}$, $\bm{61}$ and $\bm{62}$. These space groups have three glide symmetries with glide planes orthogonal to each other, three screw symmetries with the screw axes orthogonal to each other, and the inversion symmetry. Figure~\ref{ortho}(a) shows 1/8 of the Brillouin zone. Along some of the edges of the $1/8$ Brillouin zone, two glide operations with glide planes orthogonal to each other anti-commute;  this occurs only when at least one component of the wavevector $k_i\ (i=x,y,z)$ is $\pi$. Therefore, twofold or fourfold degeneracy appears along some of the edges of the 1/8 of the Brillouin zone. Emergence of the HNLs  depends on geometric arrangements of the lines with 
degeneracies. 

In Figs.~\ref{ortho}(b1)-(h1), we illustrate the geometric arrangements of the high-symmetry lines with twofold and fourfold degeneracies. Each high-symmetry line is invariant under two different glide operations. We label high-symmetry lines with twofold degeneracy as S if two Bloch states have the same eigenvalues for at least one of the glide (or mirror) operations, and as OPP otherwise. Furthermore, when the eigenvalues are the same for the glide operation with the glide plane perpendicular to the $x$ axis for example, it is labeled as ``S(x)''. 
We show high-symmetry lines with fourfold degeneracy as the red double lines, labeled as DNLs (double nodal lines). 

We take the space group No.$\bm{60}$ as an example. The label ``S(x)'' on the Z-T line means that two Bloch states have the same eigenvalues of the glide operation perpendicular to the $x$-axis. 
For other glide operations, the eigenvalues are of the opposite signs;
in this case, the two states have the opposite eigenvalues of the glide operation $G_n$ perpendicular to $z$-axis.  The label ``OPP'' on the Y-T line means that two Bloch states have the opposite eigenvalues of the glide operation $G_b$ and $G_c$, whose mirror planes are perpendicular to the $x$ and $y$ axes, respectively. From the discussion in Sec.~I\hspace{-.1em}I, an HNL appears between the Z-T line and the Y-T line on the $k_x=0$ plane, which is invariant under the glide operation $G_b$. However, HNLs do not appear on the $k_z=\pi$ plane. That is because, for the glide operation $G_n$, the case (i) does not appear on this plane . 

The resulting nodal lines in these 7 space groups are illustrated in Figs.~\ref{ortho}(b2)-(h2). We obtained these results from our theory explained so far in this paper, and we explain the derivation in Appendix C.
These HNLs have the features described in Sec.~I\hspace{-.1em}I: the appearance of these HNL is guaranteed by twofold degenerate high-symmetry lines, one with the same glide eigenvalues, and the other with the glide eigenvalues of the opposite signs. As a result of the calculation, we find that all the HNLs in these space groups cross at fourfold-degenerate points. We illustrate these fourfold-degenerate points as the red points in Figs.~\ref{ortho}(b2)-(h2). For convenience of readers, in Appendix A, we explain the mechanism of the emergence of the fourfold degeneracy in the orthorhombic space groups.

\subsection{Tetragonal primitive space groups}
\begin{figure}[t]
\centerline{\includegraphics[width=8cm,clip]{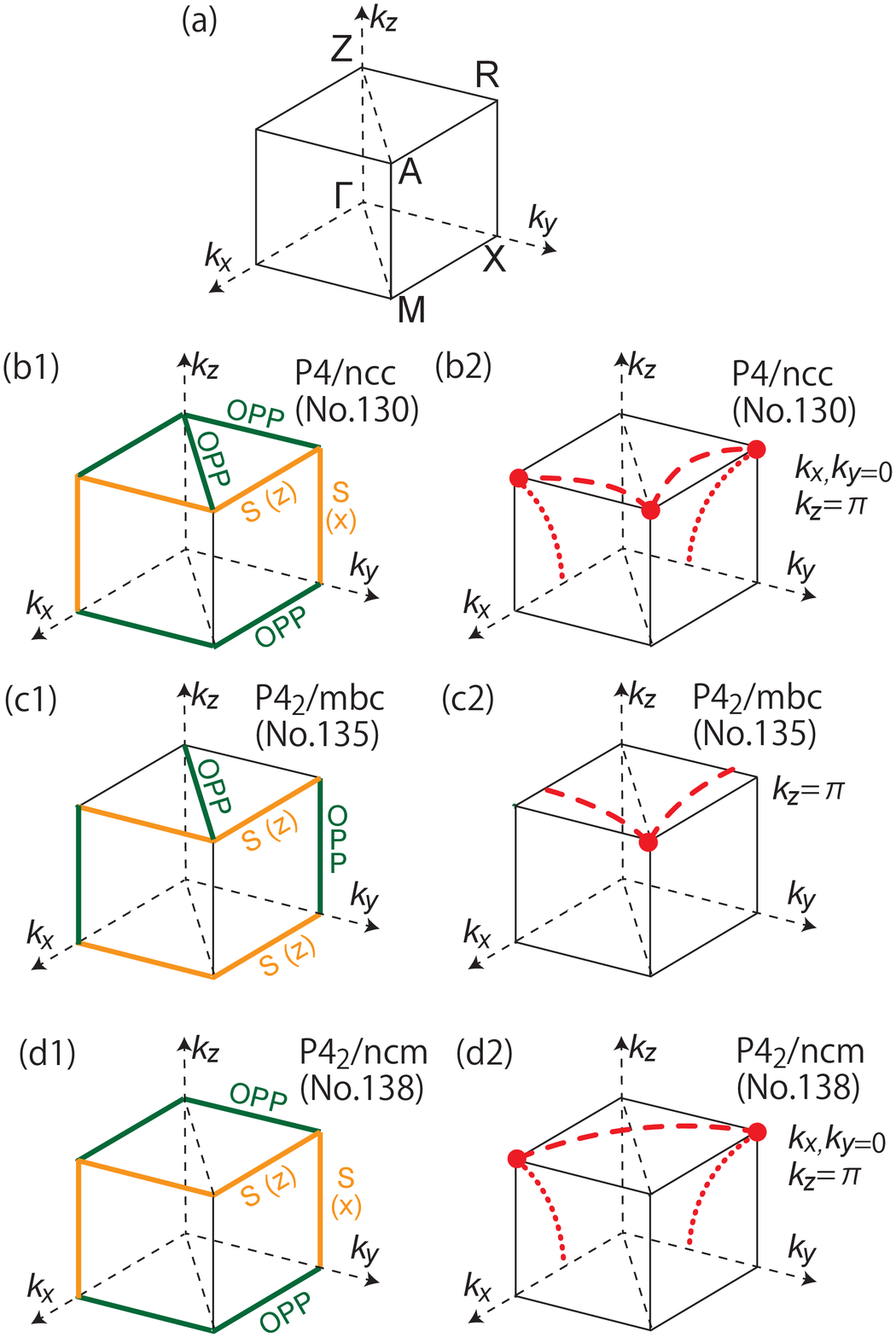}}
  \caption{(Color online) HNLs enforced by the symmetry in the tetragonal space groups No.~$\bm{130}$, $\bm{135}$ and $\bm{138}$. (a) Brillouin zone and symmetry labels for the tetragonal primitive lattice. 
(b1)-(d1) Degeneracies along the high-symmetry lines and their glide eigenvalues, which correspond to high-dimensional irreducible representations. (b2)-(h2) HNLs in the tetragonal space groups No.~$\bm{130}$, $\bm{135}$ and $\bm{138}$. The notations are the same with those in Fig.~\ref{ortho}.}
\label{tetra}
\end{figure}

In this section, we discuss the following 4 tetragonal space groups: No.~$\bm{130}$, $\bm{133}$, $\bm{135}$ and $\bm{138}$. Among them, we found that No.$\bm{133}$ do not have HNLs enforced by the symmetry. In spinless systems with the space group No.$\bm{133}$, there are several choices of irreducible representations, and some of them do not have HNLs. Therefore, general systems with the space group No.$\bm{133}$ may not have HNLs. As compared with the  orthorhombic space groups, in the tetragonal space groups, the additional fourfold screw symmetry doubles the number of symmetry operations, and additional glide-invariant planes along $(110)$ and $(1\bar{1}0)$ appear. Along the $\Gamma$-Z and M-A lines, which are invariant under the fourfold screw symmetry, there are various irreducible representations due to the screw symmetries, and one cannot conclude existence of HNLs from the space-group symmetry alone. Therefore, we focus on other high-symmetry lines. In Figs.~\ref{tetra}(b1)-(d1), we illustrate geometric arrangement of high-symmetry lines with twofold or fourfold degeneracy other than the $\Gamma$-Z and M-A lines. In Figs.~\ref{tetra}(b2)-(d2), we illustrate the resulting locations of the HNLs enforced by the symmetry. 
In Appendix B, we explain the mechanism of emergence of the fourfold degeneracy in the tetragonal primitive space groups.

\subsection{Cubic primitive space group}
\begin{figure}[t]
\centerline{\includegraphics[width=8.5cm,clip]{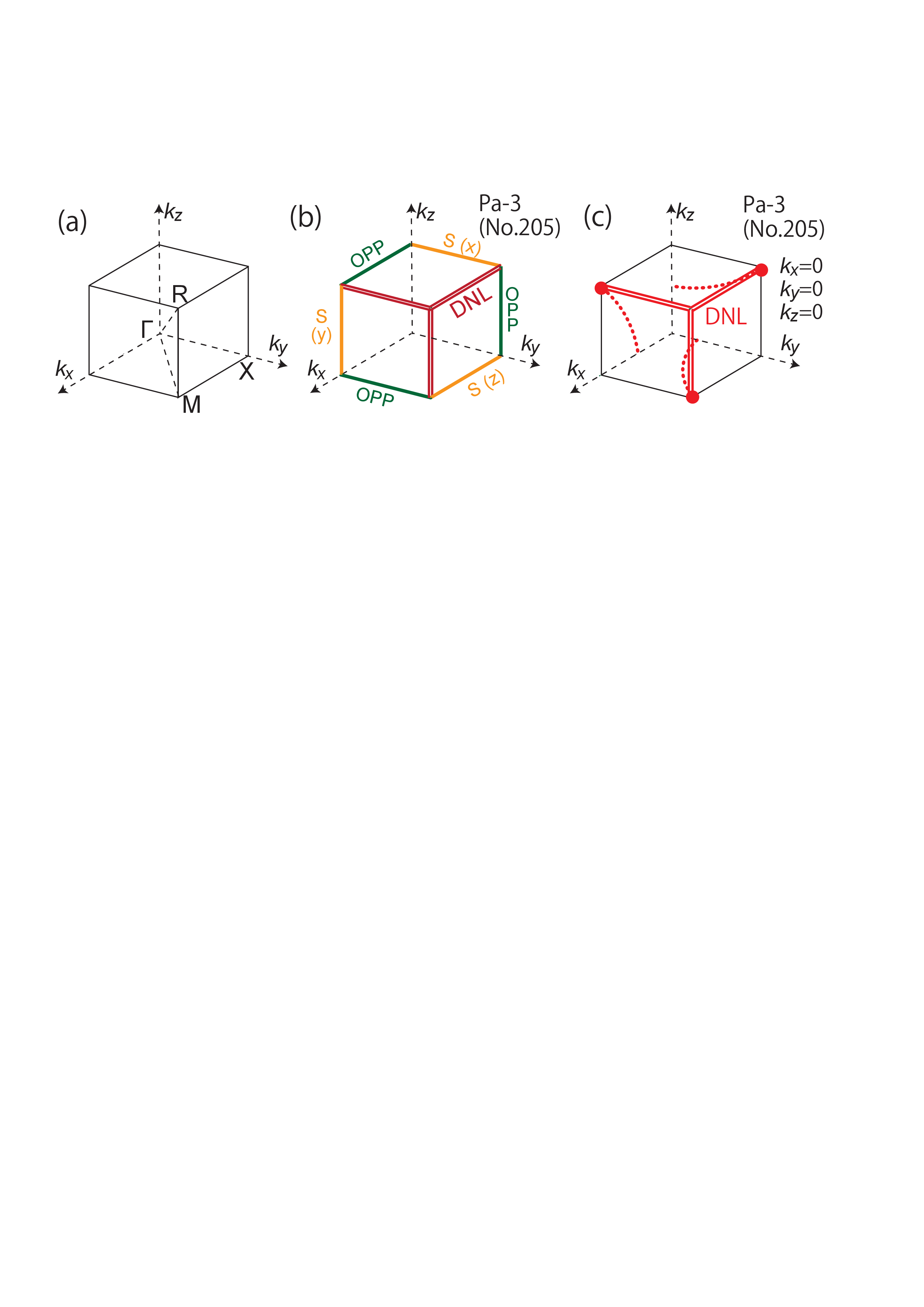}}
  \caption{(Color online) HNLs enforced by the symmetry in the cubic space group No.$\mathbf{205}$. (a) Brillouin zone and the symmetry labels for the cubic primitive lattice. (b) Degeneracy along the high-symmetry lines and their glide eigenvalues, which corresponds to high-dimensional irreducible representations. (c) HNLs in the cubic space group No.$\bm{205}$. The notations are the same with those in Fig.~\ref{ortho}.}
    \label{tri1}
\end{figure}

In this section, we discuss the cubic space group No.$\mathbf{205}$. Because the space group No.$\mathbf{61}$ is a subgroup of No.$\mathbf{205}$, HNLs in No.$\mathbf{205}$ are almost the same as those in No.$\mathbf{61}$. In Fig.~\ref{tri1}, we illustrate geometric arrangement of high-symmetry lines with twofold or fourfold degeneracy, and locations of the HNLs.  

\subsection{Non-primitive space group}
\label{sec:nonpri}
In this section, we discuss systems with the following 5 non-primitive space groups: No.$\bm{73}$, $\bm{142}$, $\bm{206}$, $\bm{228}$ and $\bm{230}$. From the same analysis with that in Sec.~I\hspace{-.1em}V A-C, we found that these space groups do not have HNLs enforced by symmetry. It is either because there are no 2-dimensional irreducible representations with the same glide eigenvalues, or because there are too many options of irreducible representations, depending on the space groups. For example, in the space group No.$\mathbf{73}$, only the three planes $k_{x,y,z}=0$ are glide-invariant planes. Therefore, no twofold degeneracy appears with the same glide eigenvalues. Other examples are No.$\mathbf{142}$ and No.$\mathbf{228}$, which have several choices of irreducible representations. Some of them enforce HNLs, but others do not. Therefore, general systems with these space group symmetries may not have HNLs.

\section{Hourglass nodal lines in Al$_3$FeSi$_2$}

\begin{figure}[t]
  \centerline{\includegraphics[width=8cm,clip]{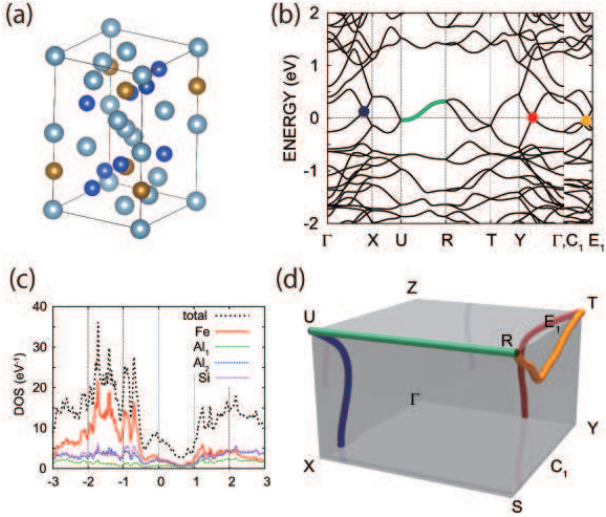}}
  \caption{(Color online) (a) Crystal structure of Al$_3$FeSi$_2$. The blue, grey, and yellow balls represent Si, Al, and Fe atoms, respectively. 
There are four Al$_1$ and eight Al$_2$ atoms in a unit cell, 
classified in terms of symmetry-equivalent positions.
Al$_1$ atoms are located at positions with high symmetry,
specifically (0,0,0),(0.5,0.5,0.5),(0,0,0.5),(0.5,0.5,0) in relative coordinates. 
(b) Electronic band structure of Al$_3$FeSi$_2$. The energy is measured from the Fermi level.
C$_1$/E$_1$ is the midpoint of the Y-S/T-R line.
(c) Total and partial densities of states of Al$_3$FeSi$_2$.
(d) Nodal lines around the Fermi level. 
The green line along the U-R line represents a double nodal line having 
fourfold degeneracy, and the others (yellow, red and blue) represent nodal lines with twofold degeneracy. The green curve, and the dots in yellow, red or blue in (b) represent the nodal lines shown in (d) in the same color. 
}
\label{AFS}
\end{figure}

Thus far, we have found space groups with the HNLs enforced by the symmetry and the filling. Therefore, to find materials with such HNLs, we should search for materials with the listed space groups with negligible spin-orbit coupling. The fillings should be $(8N+4)$ electrons per unit cell including spin degeneracy, where $N$ is an integer. Here, we propose Al$_3$FeSi$_2$ to have symmetry-enforced HNLs. 

We calculate the band structures within the density functional theory using  \textit{ab initio} code OpenMX~\cite{openmx}. 
The electronic structure is calculated in the generalized gradient approximation (GGA).
The $12\times 12\times 8$ regular ${\bf k}$-mesh including the $\Gamma$ point is employed for Al$_3$FeSi$_2$. 

Here we show that Al$_3$FeSi$_2$ is one of the materials with symmetry-enforced HNLs.
The space group of Al$_3$FeSi$_2$ is $Pbcn$ (No. $\bm{60}$) and the crystal structure is shown in Fig.~\ref{AFS}(a)~\cite{Gueneau95}. 
Figure~\ref{AFS}(b) is the electronic structure of Al$_3$FeSi$_2$.
We have checked that the system is non-magnetic in the spin-dependent calculation;
this is supported by our calculation on possible ferromagnetic and several antiferromagnetic phases, both of which lead to convergence to a nonmagnetic phase.
This nonmagnetic phase reflects a valley in the density of states near the Fermi level (Fig.~\ref{AFS}(c)). The electronic band structure near the Fermi level originates from the Fe $3d$ orbitals hybridizing with orbitals from Al and Si. 
As shown in Fig.~\ref{AFS}(a), the unit cell contains four Fe atoms surrounded by ten atoms with $p$ orbitals (Al and Si) per one Fe atom, and all the Fe atoms are related by glide operations with each other. 
As shown in Fig.~\ref{AFS}(b), the four bands are degenerate at the edge of the Brillouin zone $k_x=k_z=\pi $, i.e. the U-R line, regardless of the value of $k_y$.
This  degeneracy originates from the space-group symmetry and the time-reversal symmetry as we discussed in Sec.~I\hspace{-.1em}V. The nodal lines are shown in Fig.~\ref{AFS}(d), which perfectly agree with our prediction in Fig.~\ref{ortho}(f2).
We note that the nodal lines here have some dispersions and they are not at the Fermi energy in general. Nevertheless, we can regard this materials as a nodal-line metal within our convention.

\section{Conclusion and Discussion}
In the present paper, we showed that in spinless systems with both the inversion and the time-reversal symmetries, symmetry-enforced HNLs appear in specific space groups, when the filling factor is $4N+2$ ($N$: integer) excluding the spin degree of freedom ($8N+4$ including the spin degeneracy). We listed all the space groups with the HNLs enforced by the symmetry, and showed corresponding connectivity of nodal lines. In particular, in the spinless systems, we found that the HNLs enforced by the symmetry start from fourfold-degenerate points. These results are confirmed for the tight-binding model with the $P4_2/mbc$ (No~.$\bm{135}$) crystal symmetry and the time-reversal symmetry. We show that this model has the symmetry-enforced HNLs, regardless of the values of the parameters of the Hamiltonian. We also showed that Al$_3$FeSi$_2$ is an example of materials with the symmetry-enforced HNLs. Our results are helpful in searching candidate materials with HNLs. 

In reality, every real material has spin-orbit coupling, although it may be tiny. Such a
small spin-orbit coupling will open a gap along the nodal line. It does not mean that our theory is meaningless; our theory is useful for explaining smallness of a gap along the otherwise gapless nodal lines in such systems.

Our results can be helpful in searching candidate materials of topological insulators as well. 
That is because the $\mathbf{Z}_2$ topological number for the line nodes from Ref. \onlinecite{PhysRevLett.115.036806} is also the weak-topological insulator index. Therefore, if the line nodes are gapped in real crystals with spin-orbit coupling, the system may become
a topological insulator when the time-reversal symmetry are preserved.

Our results in the present paper 
stem from single-valued irreducible representations of the space groups. Therefore, these results can be
applied to bosonic systems, such as photons and phonons as well, 
as long as time-reversal and inversion symmetries are preserved.
Namely, in bosonic systems, symmetry-enforced HNLs are present only in the space groups $\bm{52}$, $\mathbf{54}$, $\mathbf{56}$, $\mathbf{57}$, $\mathbf{60}$, $\mathbf{61}$, $\mathbf{62}$, $\mathbf{130}$, $\mathbf{135}$, $\mathbf{138}$, $\mathbf{205}$, and the positions of the symmetry-enforced HNLs are 
shown in Figs.~\ref{ortho}, \ref{tetra}, and \ref{tri1}. 
Systems with other space group symmetries may have nodal lines but 
they are not enforced by symmetry.

There are some earlier studies on HNLs enforced by symmetry in spinful systems. In the spinful systems, because of the difference of irreducible representations, the positions of the HNLs are totally different from those in spinless systems discussed in the present paper. For example, in Ref.~\onlinecite{bzdusek2016nodal-nature}, HNLs in non-centrosymmetric spinful systems are systematically discussed. Unlike ours, their HNLs appear between a high-symmetry point and a high-symmetry line or between high-symmetry points. Therefore, the HNLs are not fixed at fourfold-degenerate points, unlike those in the spinless systems in our theory. We also note that in Ref.~\onlinecite{PhysRevB.93.155140} and Ref.~\onlinecite{chen2016topological}, HNLs in spinful systems with the space group No.$\bm{62}$ are discussed and applied to materials, SrIrO$_3$ and CaGaPt, respectively. Because SrIrO$_3$ and CaGaPt have a strong spin-orbit coupling, the positions of their HNLs are different from our theory.

Discussions of band connectivity focusing instead on topological insulating phases were recently presented in Refs. \onlinecite{kruthoff2016topological}, \onlinecite{po2017complete} and \onlinecite{bradlyn2017topological}. In the language of those papers, the HNLs presented here are features of the minimal spinless band connectivities of the space groups listed in Sec.~I\hspace{-.1em}V.

\begin{acknowledgments}
We are grateful to H.~Watanabe and M.~G.~Yamada for useful comments.
This work was supported by JSPS KAKENHI Grant Numbers 
 26287062 and 16K13834, and by the MEXT Elements Strategy Initiative to Form Core Research
Center (TIES).
\end{acknowledgments}

\appendix{}
\section{Calculation of irreducible representation for orthorhombic space groups}
From our result in Sec.~I\hspace{-.1em}I, the HNLs enforced by the symmetry always go across a high-symmetry $\bm{k}$-point with fourfold degeneracy. Here, we explain how the fourfold degeneracy in the 
orthorhombic primitive space groups appears.

We consider orthorhombic systems with the following space-group symmetries: 
\begin{align}
P&=
\left\{
\ 
\begin{pmatrix}
 -1 & & \\
 & -1 & \\
 & & -1 \\
\end{pmatrix}
\ 
\Bigg{|}
\ 
\begin{pmatrix}
0 \\
0 \\
0 \\
\end{pmatrix}\ \right\},
\notag \\
G_1&=
\left\{
\ 
\begin{pmatrix}
 -1 & & \\
 & -1 & \\
 & & 1 \\
\end{pmatrix}\ 
\Bigg{|}\,
\frac{1}{2}\mathbf{d}_1\right\},
\notag \\
G_2&=
\left\{
\ 
\begin{pmatrix}
 -1 & & \\
 & 1 & \\
 & & -1 \\
\end{pmatrix}\ 
\Bigg{|}\,
\frac{1}{2}\mathbf{d}_2\right\},
\end{align}
where $\mathbf{d}_i=\sum_{\alpha} d_i^{\alpha} \mathbf{a}^{\alpha} (i=1,2)$ ($d_{i}^{\alpha}=0,1$) is a linear combination of the primitive vectors $\mathbf{a}^{\alpha} (\alpha=x,y,z)$, and we set the lengths of the vectors $\mathbf{a}^{\alpha}$ to be unity for simplicity. $P$ represents the inversion operation. $G_1$ and $G_2$ represent $2$-fold screw rotations. Therefore $PG_1$ and $PG_2$ are glide reflections. Commutation relations among $P,G_1,G_2$ are as follows:
\begin{align}
&P^2=E,\quad G_1^2=T_z^{d_1^z},\quad G_2^2=T_y^{d_2^y},& \\
&G_1P=T_x^{d_1^x}T_y^{d_1^y}T_z^{d_1^z}PG_1,& \\
&G_2P=T_x^{d_2^x}T_y^{d_2^y}T_z^{d_2^z}PG_2,& \\
&G_1G_2=T_x^{(d_1^x-d_2^x)}T_y^{-d_2^y}T_z^{d_1^z}G_2G_1.&
\end{align}
Here, $T_{x,y,z}$ are the unit translation operators in the $x,y,z$ directions, respectively. These complicated relations become simpler at the TRIM, because $T_x,T_y,T_z=\pm1$ at the TRIM. Therefore, we only need to calculate the following $2^5=32$ patterns:
\begin{align}
&G_1^2=(-1)^{\delta_A},\ G_2^2=(-1)^{\delta_B},& \\
&G_1P=(-1)^{\delta_C}PG_1,& \\
&G_2P=(-1)^{\delta_D}PG_2,& \\
&G_1G_2=(-1)^{\delta_E}G_2G_1.&
\end{align}
Here, $\delta_{A,\cdots,E}$ are either 0 or 1, and their values are determined from the momentum $k_{x,y,z}=0,\pi$ and the values of $d_{1,2}^{x,y,z}=0,1$ as follows:  
\begin{align}
&\delta_A=\frac{k_z}{\pi}d_{1}^{z},\delta_B=\frac{k_y}{\pi}d_2^y,& \\
&\delta_C=\frac{k_x}{\pi}d_1^x+\frac{k_y}{\pi} d_1^y+\frac{k_z}{\pi} d_1^z,& \\
&\delta_D=\frac{k_x}{\pi} d_2^x+\frac{k_y}{\pi} d_2^y+\frac{k_z}{\pi} d_2^z,& \\
&\delta_E=\frac{k_x}{\pi} (d_1^x-d_2^x)-\frac{k_y}{\pi} d_2^y+\frac{k_z}{\pi} d_1^z.&
\end{align}
Once the commutation relations are given, we can calculate whether fourfold degeneracy appears at each high-symmetry point; the result is listed on Table \ref{tab:a}. 
Here, we used $\Theta^2=\mathbf{1}$ in spinless systems.

For example,  $((-1)^{\delta_A},\cdots,(-1)^{\delta_E})=(-++-+)$ at the $T$ point in the space group $Pbcn$ (No.$\mathbf{60}$). In this case, $P$ and $G_2$ anti-commute. Here, we introduce real matrices $\Gamma_1$ and $\Gamma_2$ with the following properties:
\begin{align}
\Gamma_1^2=\Gamma_2^2=1,\quad \Gamma_1\Gamma_2=-\Gamma_2\Gamma_1.
\end{align}
Then, the operators $P$, $G_1$ and $G_2$ are represented as
\begin{align}
P=\Gamma_1,\quad G_1=\pm i\bm{1},\quad G_2=\Gamma_2,
\end{align}
because $G_1$ commutes with $P$ and with $G_2$, and $G_1^2=-\bm{1}$.
 If the time-reversal symmetry is absent, we obtain two different two-dimensional irreducible representations with different signs of $G_1$. For example, they are represented by the Pauli matrices $\tau_{x,y,z}$ as follows: $P=\tauz,\ G_1=\pm i\bm{1},\ G_2=\taux$. 
In our case, the time-reversal symmetry is preserved. Because the eigenvalues of $G_1$ are pure imaginary, their signs are flipped by the time-reversal operation. Therefore, the two irreducible representations are related by time-reversal symmetry, and become a 4-dimensional co-representation. 
For example, based on the eigenstates of $P$ and $G_1$, they are represented as follows: 
\begin{align}
P=\tauz,\quad G_1=i\muz,\quad G_2=\taux,\quad \Theta=\mux K,
\end{align}
where $\mu_i$ are the Pauli matrices. 
If we include the spin degree of freedom, we use a tensor product with the spin matrices. 
In our case, the representation is given by 
\begin{align}
P=\tauz,\ G_1=i\muz\otimes i\sigmaz,\ G_2=\taux\otimes i\sigmay, \notag \\
\Theta=\mux\otimes i\sigmay K.
\end{align}
This representation is not irreducible, but can be reduced to two irreducible representations. 
In this case, $PG_1$ commutes with all the elements. Therefore, two irreducible representations are distinguished by the eigenvalues of $PG_1$. So far, we have discussed irreducible representations at high-symmetry points. Irreducible representations at high-symmetry lines are also calculated similarly; the only difference is that the number of the symmetry operations is reduced because the symmetry is lower.

\begin{table}[t]
\begin{ruledtabular}
\begin{tabular}{lcccc}
\textrm{Space group}&
\textrm{$\bm{k}$-point}&
\textrm{$((-1)^{\delta_A},\cdots,(-1)^{\delta_E})$}\\
\colrule
$\bm{52}$ \  $Pnna$ &  $S$ & $(+--+-)$ \\
\hline
$\bm{54}$ \  $Pcca$ &  $U$ & $(++---)$ \\
 &  $R$ & $(++---)$ \\
\hline
$\bm{56}$ \ $Pccn$ &  $U$ & $(++---)$ \\
 &  $T$ & $(+--+-)$ \\
\hline
$\bm{57}$ \ $Pbcm$ &  $T$ & $(---++)$ \\
 &  $R$ & $(---++)$ \\
\hline
$\bm{60}$ \ $Pbcn$ &  $U$ & $(-++-+)$ \\
 &  $T$ & $(-++-+)$ \\
 &  $R$ & $(-+--+)$ \\
\hline
$\bm{61}$ \ $Pbca$ &  $U$ & $(-++-+)$ \\
 &  $S$ & $(+---+)$ \\
 &  $T$ & $(---++)$ \\
 &  $R$ & $(--++-)$ \\
\hline
$\bm{62}$ \ $Pnma$ &  $S$ & $(+---+)$ \\
 &  $R$ & $(--+--)$ \\
\end{tabular}
\end{ruledtabular}
\caption{List of the orthorhombic space groups which have a four-dimensional irreducible representation at a high symmetry point. The first column shows the numbers and the short symbols of the space groups. The second column shows the $\bm{k}$-points with 4-fold degeneracy. The third column shows the signs of $(-1)^{\delta_i}$ ($i=A,\cdots,E$).}
\label{tab:a}
\end{table}

\section{Calculation of irreducible representations for tetragonal space groups}
We can extend our calculation in Appendix A to tetragonal space groups. In the tetragonal space groups, we have the following three symmetries: 
\begin{align}
P&=
\left\{
\ 
\begin{pmatrix}
 -1 & & \\
 & -1 & \\
 & & -1 \\
\end{pmatrix}
\ 
\Bigg{|}
\ 
\begin{pmatrix}
0 \\
0 \\
0 \\
\end{pmatrix}\ \right\},
\notag \\
\tilde{G}_1&=
\left\{
\ 
\begin{pmatrix}
 & -1 & \\
 1 & & \\
 & & 1 \\
\end{pmatrix}\ 
\Bigg{|}\,
\frac{1}{2}\mathbf{d}_1\right\},
\notag \\
G_2&=
\left\{
\ 
\begin{pmatrix}
 -1 & & \\
 & 1 & \\
 & & -1 \\
\end{pmatrix}\ 
\Bigg{|}\,
\frac{1}{2}\mathbf{d}_2\right\},
\end{align}
where $\mathbf{d}_i=\sum_{\alpha} d_i^{\alpha} \mathbf{a}^{\alpha} (i=1,2)$ ($d_{i}^{\alpha}=0,1$) is a linear combination of the Bravais lattice vector $\mathbf{a}^{\alpha} (\alpha=x,y,z)$, and $P,G_2$ are the same with the orthorhombic cases. $\tilde{G}_1$ represents the fourfold screw rotation, which is absent in the orthorhombic space groups. By calculating commutation relations between $P,\tilde{G}_1,G_2$, we can calculate irreducible representations at TRIM in the similar way as in the orthorhombic space groups.
 
Commutation relations of $P,\tilde{G}_1,G_2$ are as follows:
\begin{align}
&P^2=E,\quad \tilde{G}_1^4=T_z^{2d_1^z},\quad G_2^2=T_y^{d_2^y},& \label{tetra1} \\
&\tilde{G}_1P=T_x^{d_1^x}T_y^{d_1^y}T_z^{d_1^z}P\tilde{G}_1,& \label{tetra2} \\
&G_2P=T_x^{d_2^x}T_y^{d_2^y}T_z^{d_2^z}PG_2,& \label{tetra3} \\
&\tilde{G}_1^2P=T_x^{d_1^x-d_1^y}T_y^{d_1^x+d_1^y}T_z^{2d_1^z}P\tilde{G}_1^2,& \label{tetra4} \\
&\tilde{G}_1G_2\tilde{G}_1=T_x^{(d_1^x-d_1^y-d_2^x-d_2^y)/2}T_y^{(-d_1^x+d_1^y+d_2^x-d_2^y)/2}G_2,& \label{tetra5} \\
&\tilde{G}_1^2G_2=T_x^{d_1^x-d_1^y-d_2^x}T_y^{-d_2^y}G_2\tilde{G}_1^2.& \label{tetra6}
\end{align}

We can simplify these relations at the high-symmetry points $\Gamma,Z,M$ and $A$, because $T_x,T_y,T_z=\pm1$ and $T_x=T_y$ at these points. For the other TRIM, $R$ and $X$, we can calculate similarly as in the orthorhombic cases, because they do not have fourfold screw symmetry. 

At the high-symmetry points $\Gamma, Z, M$ and $A$, because $T_x=T_y$,  Eqs.~(\ref{tetra2})-(\ref{tetra6}) become: 
\begin{align}
&\tilde{G}_1P= T_x^{d_1^x+d_1^y}T_z^{d_1^z}P\tilde{G}_1, \label{4} \\
&G_2P= T_x^{d_2^x+d_2^y}T_z^{d_2^z}PG_2, \\
&\tilde{G}_1^2P= P\tilde{G}_1^2, \\
&\tilde{G}_1G_2\tilde{G}_1= T_x^{-d_2^y}G_2, \label{tetra7} \\
&\tilde{G}_1^2G_2= T_x^{d_1^x-d_1^y-d_2^x-d_2^y}G_2\tilde{G}_1^2. \label{2}
\end{align}
Because $T_x,T_y,T_z=\pm1$ at highest-symmetry points, the factors in the r.h.s of Eqs. (\ref{4})-(\ref{2}) involving $T_i$ are either $+1$ or $-1$. 
Therefore, $T_x^{-d_2^y}$ in Eq. (\ref{tetra7}) is equal to $G_2^2=T_y^{d_2^y}$.

From the last two equations, we obtain
\begin{align}
&\tilde{G}_1^2G_2\tilde{G}_1^2=T_x^{-2d_2^y}G_2, \\
&\tilde{G}_1^2G_2\tilde{G}_1^2=T_x^{d_1^x-d_1^y-d_2^x-d_2^y}G_2\tilde{G}_1^4=T_x^{d_1^x-d_1^y-d_2^x-d_2^y}G_2. \label{3}
\end{align}
Therefore, the factor involving $T_x$ on the right-hand side of Eq. (\ref{2}) only takes the positive sign.

We calculate the following $2^3=8$ patterns: 
\begin{align}
G_2^2= (-1)^{\delta_A} E,
\tilde{G}_1P= (-1)^{\delta_B} P\tilde{G}_1,
G_2P= (-1)^{\delta_C} PG_2.
\end{align}
where $\delta_{A,\cdots,C}$ are either 0 or 1, determined from the momentum $k_{x,y,z}=0,\pi$ and the values of $d_{1,2}^{x,y,z}=0,1$ are as follows:  
\begin{align}
&\delta_A=\frac{k_y}{\pi}d_2^y,& \\
&\delta_B=\frac{k_x}{\pi}d_1^x+\frac{k_y}{\pi} d_1^y+\frac{k_z}{\pi} d_1^z,& \\
&\delta_C=\frac{k_x}{\pi} d_2^x+\frac{k_y}{\pi} d_2^y+\frac{k_z}{\pi} d_2^z,& 
\end{align}
By a direct calculation, we can determine whether a fourfold-degenerate point appears. There are two cases: (a) only four-dimensional irreducible representations exist, or (b) in addition to four-dimensional irreducible representations, two-dimensional irreducible representations also exist. The symmetry-enforced HNLs require the case (a), as a result of the calculation for all the possible space groups. By these considerations, we obtain the list in Sec.~I\hspace{-.1em}I\hspace{-.1em}I B.

\section{Procedure for determining the positions of the HNLs in each space group}
In Figs.~\ref{ortho}, \ref{tetra}, and \ref{tri1}, we described the positions of the nodal lines derived from the irreducible representations. In this 
appendix we show how to derive these figures. We explain the procedure for the orthorhombic space groups; it is 
straightforward to extend the procedure to other space groups. 
First, we calculate irreducible representations along high-symmetry lines in $\bm{k}$-space. 
In particular, we need to calculate only on the three planes $k_i=\pi$ ($i=x,y,z$), because on other high-symmetry lines such as
the $k_x=k_y=0$ line, all the irreps are one-dimensional. 
Second, we classify these high-symmetry lines by the dimensions of irreducible representations. The case (A) represents a high-symmetry line with only two-dimensional representations. 
The case (B) represents a high-symmetry line 
with four-dimensional representations only. Lastly in the case (C) only one-dimensional irreducible representations exist, or there are some irreps with different dimensions.
We further classify the case (A) into three categories (A1)-(A3) by its glide eigenvalues. The case (A1) applies when the glide eigenvalues being the same for at least one of the glide operators. In this case we label the line with a symbol ``S $(x,y,z)$'', and show the line in yellow in the figure. Here $x$, $y$ or $z$ specifies
the direction perpendicular to the glide plane on which the glide eigenvalues are the same between the degenerated two states. 
Otherwise, when the glide eigenvalues are different for all the glide symmetries on the high-symmetry line
considered, we classify this case as (A2). It is shown in green with a label ``OPP'' in the figure. Then, if there are two types of two-dimensional irreps, one having the same glide eigenvalues and the other having the opposite signs of the glide eigenvalues, it is classified as (A3), and shown as
a black line. 
In the case (B), the states are fourfold degenerate along the high-symmetry line, meaning that it is a double nodal line. We show
this with a red double line with a label DNL in the figures. Lastly, in (C) the line is shown in black. 
In this way we obtain Figs.~\ref{ortho} (b1)-(h1), \ref{tetra} (b1)-(d1), and \ref{tri1} (b).
 
Next, from this information, we can easily obtain the positions of the  HNLs.
Here we illustrate the procedure with the example of No.~60 shown in Figs.~\ref{ortho} (f1) and (f2).
For each high-symmetry plane in $\bm{k}$-space shown in Fig.~\ref{ortho} (f1), 
if the labeling of the high-symmetry lines surrounding it falls into one of the patterns shown in 
Fig.~\ref{why-nodal-4} (a)-(f), we can draw positions of the HNLs following them. 
As an example, because the $k_y=\pi$ plane has a pattern identical with Fig.~\ref{why-nodal-4} (a), we can 
draw a nodal line on the $k_y=\pi$ plane correspondingly. By repeating this procedure for all the high-symmetry planes, 
we obtain Figs.~\ref{ortho} (b2)-(h2), \ref{tetra} (b2)-(d2), and \ref{tri1} (c).

%\bibliography{ref}

\end{document}